\documentclass{aa}
\usepackage{graphicx}
\usepackage{txfonts}
\usepackage{subfigure}
\usepackage{multirow}
\usepackage{natbib}
\usepackage{array}
\usepackage{soul}
\usepackage[normalem]{ulem}

\usepackage{lscape}
\begin{document}
   \title{ISM composition through X-ray spectroscopy of LMXBs}

%    \subtitle{IX. The Galactic foreground}

   \author{C. Pinto
          \inst{1,2,3}
          \and
%           G.~A.~Kriss \inst{2,3}
          J.~S.~Kaastra\inst{1,4}
%           F.~Verbunt\inst{1,2}
          \and
          E.~Costantini\inst{1}
          \and
          C.~de~Vries\inst{1}
          }

   \institute{SRON Netherlands Institute for Space Research,
              Sorbonnelaan 2, 3584 CA Utrecht, The Netherlands.
%               \email{c.pinto@sron.nl}
         \and
             Department of Astrophysics/IMAPP, Radboud University, PO BOX 9010, 6500 GL Nijmegen, 
             The Netherlands.
         \and
             Institute of Astronomy, Madingley Road, CB3 0HA Cambridge, 
             United Kingdom, \email{cpinto@ast.cam.ac.uk}.
         \and
             Astronomical Institute, Utrecht University,
             P.O. Box 80000, 3508 TA Utrecht, The Netherlands.\\
         }
   \date{Received 2 October 2012 / Accepted 7 January 2013}

% \abstract{}{}{}{}{} 
% 5 {} token are mandatory
 
  \abstract
   {The diffuse interstellar medium (ISM) is an integral part of the evolution of the entire Galaxy. Metals are produced by stars and their abundances are the direct testimony of the history of stellar evolution. However, the interstellar dust composition is not well known and the total abundances are yet to be accurately determined.}
   {We probe ISM dust composition, total abundances, and abundance gradients through the study of interstellar absorption features in the high-resolution X-ray spectra of Galactic low-mass X-ray binaries (LMXBs).}
   {We use high-quality grating spectra of nine LMXBs taken with XMM-Newton. We measure the column densities of O, Ne, Mg, and Fe with an empirical model and estimate the Galactic abundance gradients.}
   {The column densities of the neutral gas species are in agreement with those found in the literature. {Solids are a significant reservoir of metals like oxygen and iron.} Respectively, 15--25\% and 65--90\% of the total amount of \ion{O}{i} and \ion{Fe}{i} is found in dust. The dust amount and mixture seem to be consistent along all the lines-of-sight (LOS). Our estimates of abundance gradients and predictions of local interstellar abundances are in agreement with those measured at longer wavelengths.}
   {Our work shows that X-ray spectroscopy is a very powerful method to probe the ISM. For instance, on a large scale the ISM appears to be chemically homogeneous showing similar gas ionization ratios and dust mixtures. {The agreement between the abundances of the ISM and the stellar objects suggests that the local Galaxy is also chemically homogeneous.}}


   \maketitle
%
%________________________________________________________________

\section{Introduction}
\label{sec:introduction}

The interstellar medium (ISM) is one of the most important component of galaxies because it influences their evolution: the ISM is enriched with heavy elements by stellar evolution, and it provides the source of material for following generation of stars. There are several physical processes that alter the metallicity of the ISM. Stellar winds and supernovae expel part of the interstellar gas out of the Galactic disk, but gravity generally forces the gas to fall back through the process known as ''Galactic fountain'' \citep{Shapiro1976}. Gas accreted from smaller galaxies, like the Magellanic Clouds, and the intergalactic medium increases the reservoir of low metallicity gas. The ISM also shows a complex structure consisting of phases at different equilibrium temperatures \citep[for a review, see][]{Draine2011}. The cold phase is a blend of dust, molecules and almost neutral gas below $10^4$ K. The warm and hot phases are mostly gaseous. The warm gas is weakly ionized, with a temperature of $\sim 10^4$ K. 
The hot gas is highly ionized, with temperatures of about $10^{6}$ K. These three main phases are not entirely separated from each other. For instance, it is thought that a conductive cooling layer, revealed by \ion{O}{vi}, lies between the warm and hot ionized phases \citep[see e.g.][]{Richter2006}.

In the spectra of background sources the ISM produces reddening and absorption lines. High resolution X-ray spectroscopy has become a powerful diagnostic tool for constraining the chemical and physical properties of the ISM. The K-shell transitions of carbon, nitrogen, oxygen, neon, and magnesium, and the L-shell transitions of iron fall in the soft X-ray energy band. The presence of different ionization states constrains the multi-phase temperature structure of the ISM. A summary of the results obtained by X-ray spectroscopy on the ISM in the last decades is given by \citet{Pinto2010}.

Briefly, \citet{Schattenburg} first constrained interstellar oxygen in the X-ray band using the \textit{Einstein Observatory}, but a thorough study of the ISM was possible only after the launch of the {\textit{XMM-Newton}} and \textit{Chandra} satellites, provided with high-resolution gratings. Absorption structure due to ionized gas and dust was found around the interstellar oxygen K-shell absorption edge in the spectra of several sources \citep{paerels, devries, JuettI, costantini2005, Costantini2012, costantini, kaastra09, Pinto2010, Pinto2012, Pinto2012b}.

{However, in the last decades, the most important discoveries on the physics of the ISM phases have been obtained with UV, IR, optical, and radio spectroscopy. Dust is usually studied in IR, molecules in radio and IR, while the neutral and warm phases of the interstellar gas are also probed with UV data (see \citealt{ferriere} and references therein). The hot gas is the only ISM phase which is more often studied in X-rays.} However, despite the very high resolution and signal-to-noise ratio at long wavelengths, both the chemistry and physics of the interstellar medium are still under debate. Elemental abundances, dust depletion factors and dust chemistry are not yet well determined and thus provide an open and interesting research field. The heavy elements such as oxygen (the most abundant one) and iron are produced in high-mass stars. Molecules and dust compounds are mainly formed in AGB stars and then grow in the diffuse ISM \citep[see e.g.][]{Mattsson2012}. Heavy elements directly witness the history of 
the past stellar evolution. In the last years there have been many attempts to measure the interstellar abundances, but in many cases only lower limits were obtained due to the depletion of the elements from the gaseous phases into dust grains and to strong line saturation at long wavelengths. The dust depletion factors and especially the compounds mixture are not yet well known and thus the total ISM abundances are affected by strong systematic uncertainties. For instance, a significant fraction of the oxygen which is bound in the solid phase is still unknown: silicates and carbonaceous compounds are not abundant enough to cover the amount of depleted oxygen. Ices may solve this problem, but they are hard to detect \citep[for a review about depletion factors in the ISM see][]{Jenkins2009}. Therefore, our work focuses on an alternative method to determine ISM column densities, dust depletion factors, molecular compounds, total abundances (gas + dust), and abundance gradients through X-ray spectroscopy.

In this work we report the detection and modeling of interstellar absorption lines and edges in the high-quality spectra of nine LMXBs: \object{GS 1826$-$238}, \object{GX 9+9}, \object{GX 339$-$4}, \object{Aql X$-$1}, \object{SAX J1808.4$-$3658}, \object{Ser X$-$1}, \object{4U 1254$-$690}, \object{4U 1636$-$536}, and \object{4U 1735$-$444}. The observations are taken with the \textit{XMM-Newton} Reflection Grating Spectrometer \citep[RGS, ][]{denherder01}, see Table~\ref{table:1} for details. These sources are well suited for the analysis of the ISM because of their brightness and column densities $N_{\rm H} \sim 1-5 \times 10^{25} \,  {\rm m}^{-2}$ (see Table~\ref{table:fit_complete}), which are high enough to produce prominent O, Fe, and Ne absorption edges in the soft X-ray energy band. This paper provides a significant extension of the previous work on the ISM in the line-of-sight (LOS) towards GS 1826$-$238 done by \citet{Pinto2010}. We have used updated atomic and molecular database and extended the 
analysis to several LOS (see Fig.~\ref{fig:map_sources}).

Our analysis focuses on the $7-38$ {\AA} first order spectra of the RGS detectors. We have not used the EPIC spectra of the observations because they have much lower spectral resolution and we were not interested in determining the source continuum at energies higher than the RGS band. Most of interstellar X-ray features are indeed found in the soft X-ray energy band. We performed the spectral analysis with SPEX\footnote{www.sron.nl/spex} version 2.03.03 \citep{kaastraspex}. We scaled the elemental abundances to the proto-Solar abundances of \citet{Lodders09}. We use the $\chi^2$-statistic throughout the paper and adopt $1 \sigma$ errors.

The paper is organized as follows. In Sect.~\ref{sec:data} we present the data. In Sect.~\ref{sec:spectra} we report the spectral features that we analyze. In Sect.~\ref{sec:analysis} we describe the models we use and the results of our analysis. The discussion and the comparison with previous work are given in Sect.~\ref{sec:discussion}. Conclusions are reported in Sect.~\ref{sec:conclusion}.

\section{The data}
\label{sec:data}

\begin{figure}
      \centering
      \includegraphics[width=6.5cm, bb=55 55 540 700, angle=90]{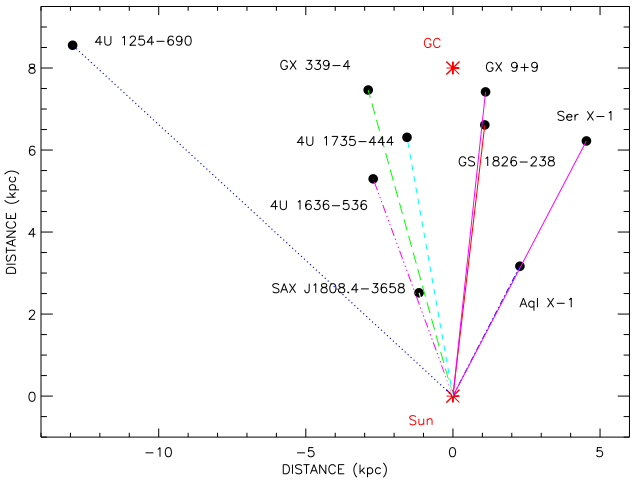}%}
      \caption{Map of the X-ray sources as projected on the Galactic plane. The \object{Sun} is assumed to be 8\,kpc far away from the \object{Galactic Center} (GC).}
          \label{fig:map_sources}
    \end{figure}

For this work we have used 22 archival spectra of nine sources taken with \textit{XMM-Newton} (see Fig.~\ref{fig:map_sources} and  Table~\ref{table:1}). These spectra were taken in periods of high-flux state of the sources and each single spectrum shows high statistics. We have excluded in our analysis additional observations taken during low-flux states. For some sources we have just one good spectrum, while for other ones we obtained up to five high-quality spectra. The spectra taken on the same source have been fitted together in order to increase the statistics, but we have used a different approach between stable and variable sources. When the persistent emission of a certain source was steady and the spectra were perfectly superimposable, we have stacked the spectra according to the procedure described by \citet{Kaastra11b}. Briefly, we have reduced the spectra with the Science Analysis System (SAS) version 11.0, obtained fluxed spectra for RGS 1 and 2 and stacked them with the SPEX task {\sl RGS\_
fluxcombine}. We cannot stack the spectra of variable sources because this may introduce spurious features near the interstellar edges. Thus, we simultaneously fit their spectra coupling the parameters of the interstellar absorbers, but adopting a different continuum for each observation. This can be done with the SPEX task {\sl sectors}.

{The LMXBs \object{GS~1826$-$238}, \object{GX~9+9}, \object{Ser~X$-$1}, \object{4U~1254$-$690}, and \object{4U~1636$-$536} showed little or insignificant spectral variability, thus we could stack their RGS~1 and 2 spectra, which were taken during all the observations, and produce a single final fluxed spectrum for each LMXB (see Figs.~\ref{fig:neon} -- \ref{fig:oxygen}). We adopted a similar approach in \citet{Pinto2010} for \object{GS~1826$-$238}. There we filtered out the bursting emission from the spectra but also reported that it was negligible compared to the persistent emission. Therefore we do not remove the periods of bursts in any of our data.}

For \object{Aql X$-$1}, \object{SAX J1808.4$-$3658}, and \object{4U 1735$-$444} we have only found one very good spectrum. Their other observations available in the archive have either too short exposure or much lower flux in such a way that they do not significantly improve the statistics. For each of these sources we simply fit the RGS 1 and 2 count spectra simultaneously. The LMXB \object{GX 339$-$4} is an exception. Its spectral continuum and slope strongly vary, but the three observations with IDs 0148 and 0605 have superimposable spectra, as well as the remaining two observations (IDs 0204). Therefore we split the spectra in these two groups and stack them producing two final fluxed spectra with comparable statistics. In Figs.~\ref{fig:neon} -- \ref{fig:oxygen} with show the Ne, Fe, and O absorption edges of the \object{GX 339$-$4} fluxed spectrum obtained by stacking the group of observations with ID 0204.

In Table \ref{table:1} we report the RGS exposure times and identification number for each observation. We also provide the Galactic coordinates, distance $d$ and hydrogen column density $N_{\rm H}$ for each source. Most of the distances have been taken from \citet{Galloway2008}, which estimated them through the luminosity peak during the LMXBs burst. The $N_{\rm H}$ ranges refer to the spread between the values measured by the Leiden/Argentine/Bonn Survey of Galactic \ion{H}{i} and the Dickey \& Lockman \ion{H}{i} in the Galaxy, see \citet{Kalberla2005} and \citet{Dickey1990}. The $N_{\rm H}$ values as provided by these two surveys differ by up to 30\% and we will determine them from the X-ray spectrum. In the RGS fits we rebin the spectra by a factor of two, i.e. about 1/3\,FWHM (the first order RGS spectra provide a resolution of 0.06$-$0.07\,{\AA}). This gives the optimal binning for the RGS and a bin size of about 0.02\,{\AA}.

\begin{table*}
\caption{\textit{XMM-Newton}/RGS observations used in this paper.}  
\label{table:1}    
\renewcommand{\arraystretch}{1.3}
\begin{center}
 \small\addtolength{\tabcolsep}{+2pt}
\scalebox{1}{%
\begin{tabular}{c c c c c c c }     
\hline\hline            
Source                       &  ID $^{(a)}$ & Length (ks) $^{(b)}$ & $l,b$ $^{(c)}$      & d (kpc) $^{(d)}$     & $N_{\rm H}$ ($10^{25}\,{\rm m}^{-2}$) $^{(e)}$\\  
%                              &              &             &                              &                      & \\
\hline   
\hline   
\multirow{3}{*}{4U 1254--690}&   0060740901 &     28.5   & \multirow{3}{*}{$303.5,-6.4$} & \multirow{3}{*}{15.5 $^{(f)}$} & \multirow{3}{*}{2.2--2.9}\\
                             &   0405510301 &     60.6   &                               &                      &   \\
                             &   0405510501 &     60.7   &                               &                      &   \\
\hline                
\multirow{4}{*}{4U 1636--536}&   0105470401 &     20.6   & \multirow{4}{*}{$332.9,-4.8$} & \multirow{4}{*}{5.95 $^{(f)}$} & \multirow{4}{*}{2.6--3.6}\\
                             &   0303250201 &     30.9   &                               &                      &   \\
                             &   0500350301 &     31.5   &                               &                      &   \\
                             &   0606070201 &     28.8   &                               &                      &   \\
\hline                
\multirow{1}{*}{4U 1735--444} &  0090340201 &     20.6   & \multirow{1}{*}{$346.1,-7.0$} & \multirow{1}{*}{6.5 $^{(f)}$} & \multirow{1}{*}{2.6--3.0}\\
\hline                
\multirow{1}{*}{Aql X--1}    &   0406700201 &     52.6   & \multirow{1}{*}{$35.7,-4.1$}  & \multirow{1}{*}{3.9 $^{(f)}$} & \multirow{1}{*}{2.8--3.4}\\
\hline                
\multirow{2}{*}{GS 1826--238}&   0150390101 &    106.3   & \multirow{2}{*}{$  9.3,-6.1$} & \multirow{2}{*}{6.7 $^{(f)}$} & \multirow{2}{*}{1.7--1.9}\\
                             &   0150390301 &     91.5   &                               &                      &   \\
\hline                
\multirow{5}{*}{GX 339--4}   &   0148220201 &     20.3   & \multirow{5}{*}{$338.9,-4.3$} & \multirow{5}{*}{8 $^{(g)}$} & \multirow{5}{*}{3.7--5.3}\\
                             &   0148220301 &     13.7   &                               &                      &   \\
                             &   0204730201 &     13.3   &                               &                      &   \\
                             &   0204730301 &     13.4   &                               &                      &   \\
                             &   0605610201 &     32.9   &                               &                      &   \\
\hline                
\multirow{2}{*}{GX 9+9}      &   0090340101 &     10.6   & \multirow{2}{*}{$8.5,9.0$}    & \multirow{2}{*}{7.5 $^{(h)}$} & \multirow{2}{*}{2.0--2.1}\\
                             &   0090340601 &     21.7   &                               &                      &   \\
\hline                
\multirow{1}{*}{SAX J1808.4--3658} & 0560180601 & 63.7   & \multirow{1}{*}{$335.4,-8.1$} & \multirow{1}{*}{2.77 $^{(f)}$} & \multirow{1}{*}{1.1--1.3}\\
\hline                
\multirow{3}{*}{Ser X--1}    &   0084020401 &     21.5   & \multirow{3}{*}{$36.1,4.8$}   & \multirow{3}{*}{7.7 $^{(f)}$} & \multirow{3}{*}{4.0--4.7}\\
                             &   0084020501 &     21.7   &                               &                      &   \\
                             &   0084020601 &     21.8   &                               &                      &   \\
\hline                
\end{tabular}}
\end{center}
$^{(a, \, b)}$ Identification number and duration of the observation. $^{(c, \, d, \, e)}$ Source Galactic coordinates in degrees, distance from the Sun, and hydrogen column density (see \citealt{Kalberla2005} and \citealt{Dickey1990}).
$^{(f)}$ Distance estimates are taken from \citet{Galloway2008}. $^{(g)}$ For GX\,339-4 we adopt the distance as suggested by \citet{Zdziarski2004}. $^{(h)}$ The distance for GX\,9+9 is the mean value between those ones provided by \citet{ChristianSwank1997} and \citet{Zdziarski2004}.\\

\end{table*}

\section{Spectral features}
\label{sec:spectra}

\begin{figure*}
\begin{center}
      \subfigure{ 
      \includegraphics[width=15cm, angle=-90]{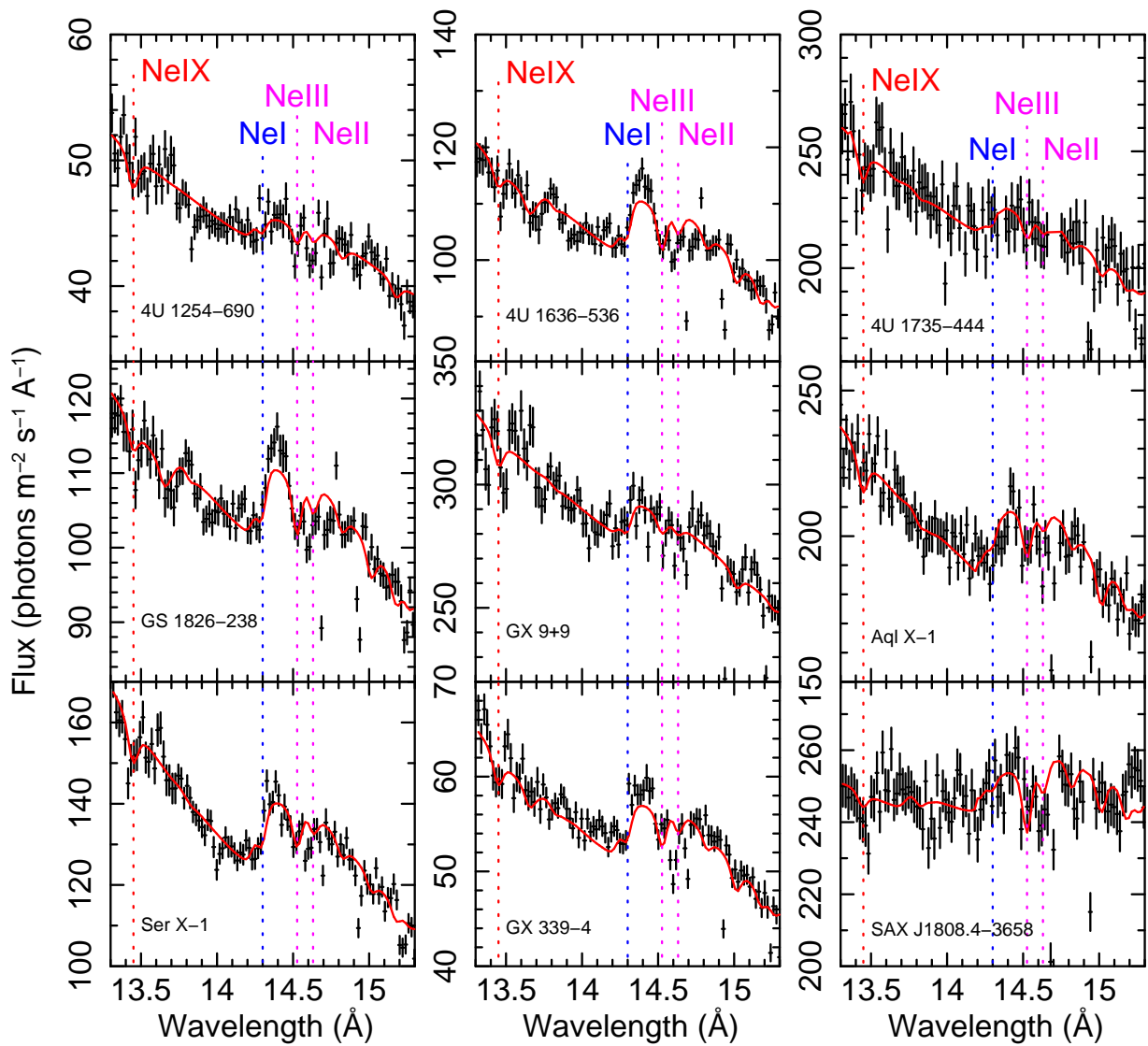}}
      \caption{Data and best-fitting models: the Ne K edge.}
          \label{fig:neon}
   \end{center}
\end{figure*}

\begin{figure*}
  \begin{center}
      \subfigure{ 
      \includegraphics[width=15cm, angle=-90]{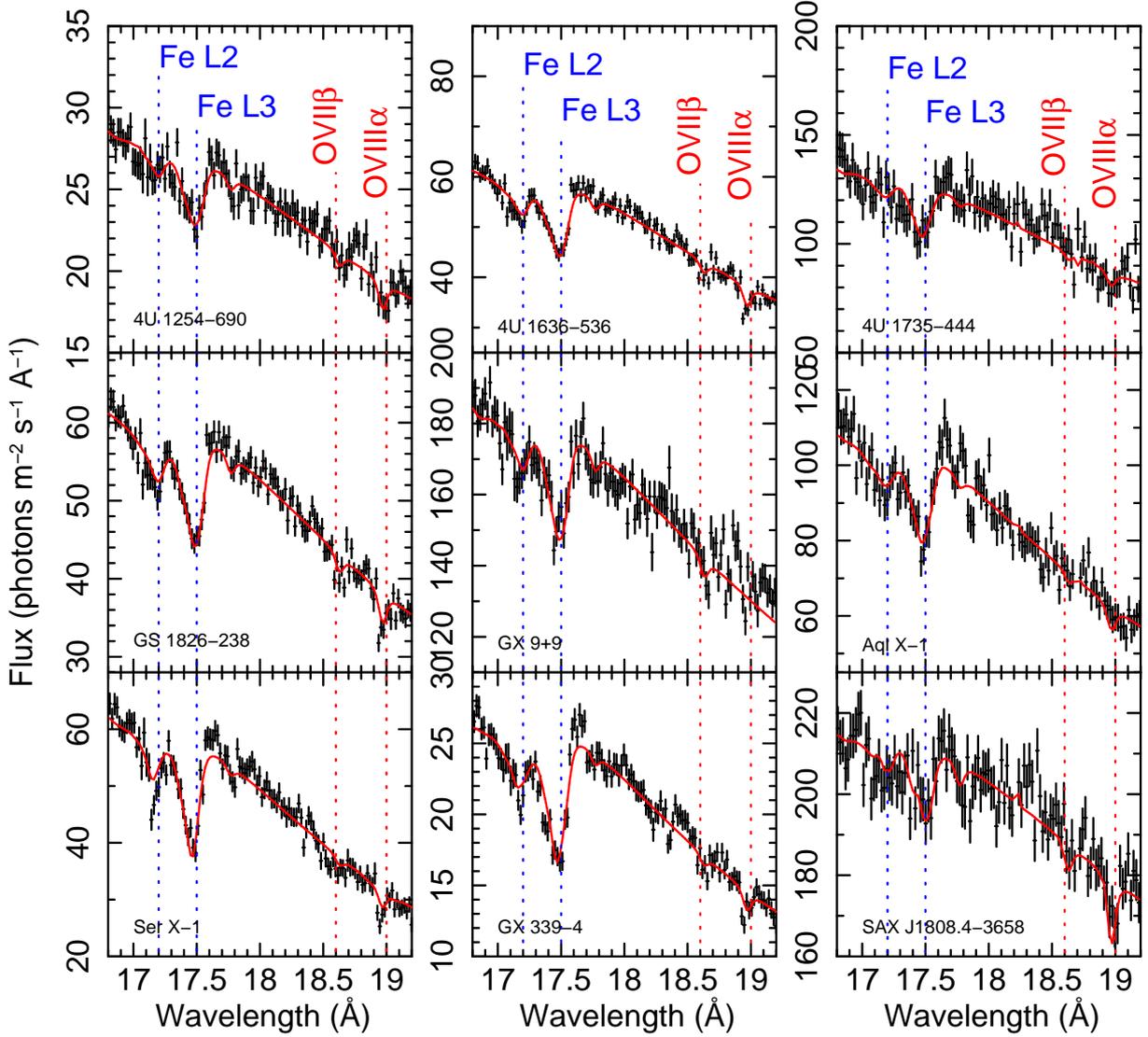}}
      \caption{Data and best-fitting models: the Fe L2 and L3 edge.}
          \label{fig:iron}
  \end{center}
\end{figure*}

\begin{figure*}
  \begin{center}
      \subfigure{ 
      \includegraphics[width=15cm, bb=30 60 560 630, angle=-90]{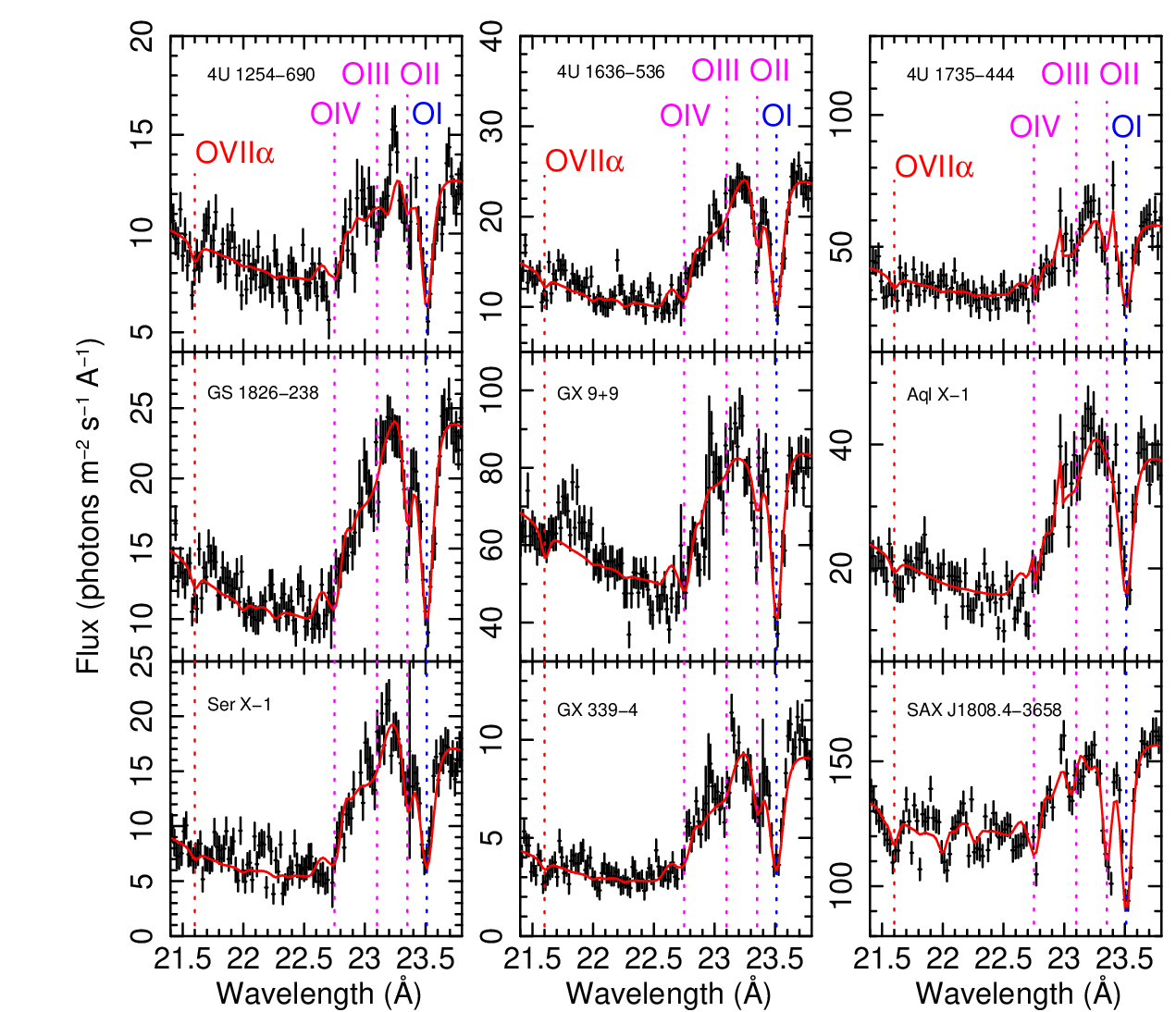}}
      \caption{Data and best-fitting models: the O K edge.}
          \label{fig:oxygen}
   \end{center}
\end{figure*}

In Figs.~\ref{fig:neon} -- \ref{fig:oxygen} we show the neutral absorption edges of neon, iron, and oxygen for the nine LMXBs, which are the strongest spectral features. SAX J1808.4--3658 also shows a prominent \ion{N}{i} 1s--2p edge at 31.3\,{\AA}, but most of the sources have weak nitrogen and magnesium edges, which means that the Mg and N abundances are not well constrained. The \ion{Ne}{i} edge (see 14.3\,{\AA} in Fig.~\ref{fig:neon}) is shallower than those of Fe and O, but it is interesting for the presence of additional absorption lines due to mildly ionized \ion{Ne}{ii-iii} and heavily ionized \ion{Ne}{ix} gas. For a few sources we are even able to detect a weak \ion{Ne}{viii} line at about 13.7\,{\AA}. {For some spectra near the neon edge two weak absorption features at 15.0\,{\AA} and 14.2\,{\AA} are visible that are presumably due to \ion{Fe}{xvii} and \ion{Fe}{xviii} in the hot gas phase.}

The iron L2 and L3 edges are located at 17.15 and 17.5\,{\AA} (see Fig.~\ref{fig:iron}). The sources show clear differences in the depth of these edges. We also provide the position of the \ion{O}{vii}\,$\beta$ (1s--3p) and \ion{O}{viii}\,$\alpha$ (1s--2p) absorption lines at 18.6 and 19.0\,{\AA}, respectively. These two lines also trace the hot ionized gas. Fig.~\ref{fig:oxygen} shows the most interesting part of the spectrum: the oxygen K edge. The \ion{O}{i} 1s--2p line at about 23.5\,{\AA} is the strongest ISM spectral feature, the neutral oxygen is also responsible for the jump in spectrum between 22.5 and 23.2\,{\AA}. Additional but not less important 1s--2p absorption lines are produced by \ion{O}{ii} (23.35\,{\AA}), \ion{O}{iii} (23.1\,{\AA}), \ion{O}{vii} (21.6\,{\AA}), and occasionally \ion{O}{vi} (22.0\,{\AA}). Dust oxygen compounds affect mainly the spectral curvature between $22.7-23.0$\,{\AA} \citep[see e.g.][and references therein]{Pinto2010}. An easy way to check for strong amounts of dust is 
to compare the depth of the edge with that of the line. Dust usually is responsible for a deep edge, while the gas also provides strong lines. For instance, in Ser~X-1 the O~K edge is deeper than the 1s--2p line, while in SAX J1808.4--3658 the absorption line is clearly deeper than the edge. This means that along the LOS towards Ser~X-1 the amount of oxygen found in dust compounds is expected to be larger. We will confirm this in Sect.~\ref{sec:discussion}. Dust is also expected to provide most of iron \citep[see e.g.][]{Jenkins2009}, which means that the bulk of the absorption between 17--17.5\,{\AA} is produced by iron dust compounds \citep[see also][]{Costantini2012}. Dust is also responsible for most of the magnesium edge at about 9.5\,{\AA}.


\section{Spectral modeling of the ISM}
\label{sec:analysis}

The advantage of using LMXBs for the analysis of the ISM lies in their {high-statistics} and simple spectrum. All the sources in our sample provided us with well exposed spectra, which are easy to model by adopting a simple continuum. In most of the cases one power-law (PL) continuum gave an acceptable and quickly-converging fit. When a single PL did not provide a good fit then we were able to obtain an acceptable fit by simply adding another PL. The continuum of LMXBs usually originates from their accretion disk and corona in the form of blackbody and comptonized emission, respectively. However, we are using only the soft part of the X-ray spectrum which is relevant for the ISM absorption and is not broad enough to provide accurate constraints on the X-ray emission components. Moreover, this work does not focus on the nature of the continuum intrinsic to the sources. Thus, we prefer to use PL components rather than physical ones like blackbody (BB) or comptonization (COMT). In principle the choice of the 
continuum component may affect the estimate of the hydrogen column density, while the other ionic columns are well constrained as they are estimated through narrow absorption lines which do not depend on the continuum. Thus, we also tested both PL and a combination of BB~/~COMT in modeling our RGS spectra {({\sl bb} and {\sl comt} models in SPEX)}. The best fit values of $N_{\rm H}$ were consistent within the errors, which validates our choice of a simple PL continuum.

We have built an empirical model for the ISM similar to that one successfully used by \citet{Pinto2012} but with a few useful changes. Essentially, they used three components modeled with the {\sl slab} model of SPEX to fit the three main phases of the interstellar gas. {The {\sl slab} model gives the transmission through a layer of gas with arbitrary ionic column densities. Here we also use two {\sl slab} components to model the warm and hot phases of the ISM, but} we prefer to model the cold neutral gas with the {\sl hot} component in SPEX. The {\sl hot} model calculates the transmission of a collisionally-ionized equilibrium plasma. For a given temperature and set of abundances, the model calculates the ionization balance and then determines all the ionic column densities by scaling to the prescribed total hydrogen column density. At low temperatures this model mimics very well the neutral interstellar gas \citep[see SPEX manual and][]{kaastra09}. Free parameters in the {\sl hot} model are the hydrogen 
column density $N_{\rm H}$, the temperature $T$, the velocity dispersion $\sigma_{\rm v}$, the systematic velocity v, and the abundances. However, in X-rays the spectral resolution is not high enough to constrain the velocities with sufficient accuracy, thus we assume that the interstellar gas is at rest. We also assume a temperature of 0.5\,eV (about 5800\,K), the lowest value available in the SPEX {\sl hot} model, because in this way the gas is mostly neutral. In X-rays with the current satellites it is not possible to resolve the narrow lines of the interstellar medium. A spectral fit can easily provide a wrong value of velocity dispersion $\sigma_{\rm v}$ and consequently of column density because they are degenerate. Therefore it is more appropriate adopting a physically reasonable value for the $\sigma_{\rm v}$ and just fitting the column density of each ion. We adopt a nominal value of 10\,km\,s$^{-1}$ for the velocity dispersion of the cold (\ion{O}{i}) and warm (\ion{O}{ii-v}) gas components, which 
is similar to that found by \citet{Pinto2012} using a higher-resolution UV spectrum. For the hot gas (\ion{O}{vi} and higher ionization states) we adopt $\sigma_{\rm v}=$100\,km\,s$^{-1}$, which is an average value between those suggested in the literature (see for instance the work by \citealt{Yao2005} and \citealt{yao09} on high-resolution \textit{Chandra} X-ray spectra of several LMXBs). We took into account absorption by interstellar dust with the SPEX {\sl amol} component. The {\sl amol} model calculates the transmission of various molecules, for details see \citet{Pinto2010}, \citet{Costantini2012} and the SPEX manual (Sect.~3.3). The model currently takes into account the modified edge and line structure around the O and Si K-edge, and the Fe K / L-edges, using measured cross sections of various compounds taken from the literature. Relevant and abundant silicates, ices and organic molecules are present in the our database. In fitting the molecular models we limit the range of column densities to 
physical values. For instance, we took care that the column density of each weakly-abundant element did not exceed the proto-Solar value predicted by the best-fitting $N_{\rm H}$. This is important especially for Ca and Al (contained respectively in andradite and hercinite), whose abundances are much lower than oxygen.

\subsection{{The fitting procedure}}
\label{sec:procedure}

{The first step is determining a preliminary model for the whole spectrum. This requires to fit the spectrum of each source with a simple model consisting of one (or more) power-law component(s) absorbed by neutral gas ({\sl hot} model in SPEX). As already mentioned, one or two power-law components are sufficient to reproduce all the spectral continua. This fit provides preliminary values for the normalization and slope of the power-law as well as the hydrogen column density and the abundances of the neutral gas. The second step is determining the contributions of each absorbing component. Some compounds and ions produce relevant absorption features only near certain absorption edges. For instance, ions such as \ion{O}{ii-vi} and molecules like water ice, andradite and pyroxene mostly affect the spectral range near the O K-edge. Therefore, we first fit each edge with the preliminary model and add in sequence ions and, in a second instance, molecular compounds looking for significant changes in the $\
chi^2$. In the case of the oxygen K-edge, significant changes in the $\chi^2$ are given by ions like \ion{O}{ii-iv}, \ion{O}{vii-viii}, and compounds like water ice, pyroxene, magnetite, and hematite. Once we have measured the $\Delta\,\chi^2$ of the absorbing components in each edge, we jointly fit the edges or the spectral ranges that share transitions from same ions or compounds. The wavelength interval between 15 and 24\,{\AA} contains the strong Fe and O edges, and the bulk of the dust signatures. The fit of this spectral region provides a starting model for the final fit extended to the whole spectrum.}

\subsection{Step-by-step analysis for GX 339-4}
\label{sec:GX3394}

\begin{figure}
      \centering
      \includegraphics[width=6.5cm, bb=75 20 585 715, angle=-90]{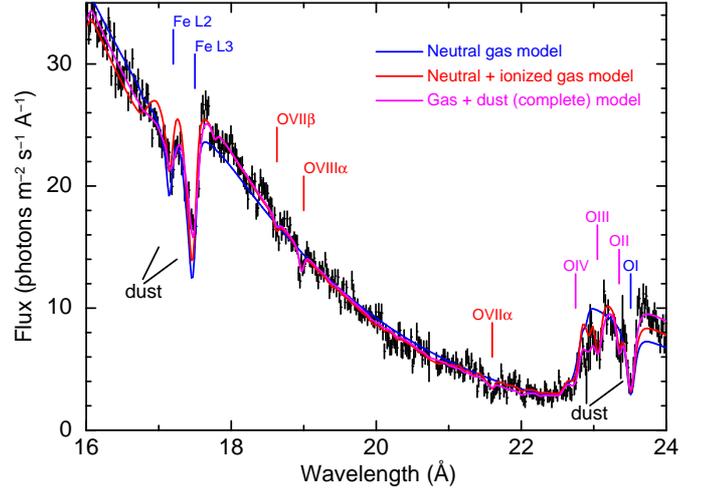}%}
      \caption{{Spectral fits of the Fe L and O K edges for GX~339-4 (for clarity only one stacked spectrum is shown).
               For more detail see Sect.~\ref{sec:GX3394} and Table~\ref{table:GX339}.}}
          \label{fig:GX339_edges}
    \end{figure}

\begin{figure}
      \centering
      \includegraphics[width=6.5cm, bb=60 85 540 700, angle=-90]{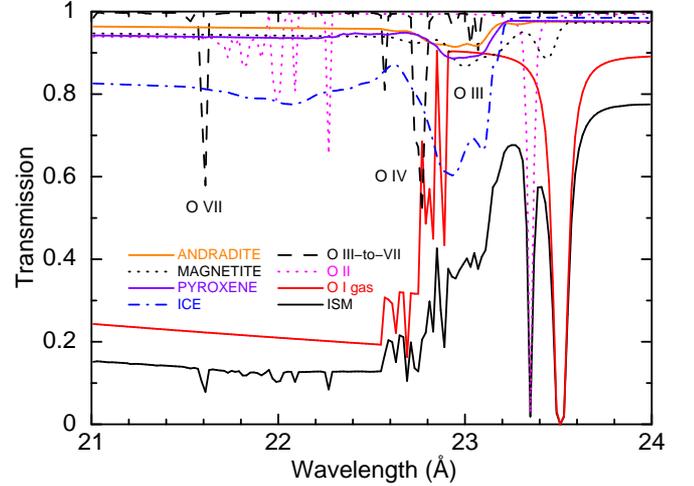}%}
      \caption{{Transmission of the ISM near the O K-edge in the spectrum of GX~339-4.
               The model refers to the best-fitting model as given in Table~\ref{table:fit_complete}. 
               The solid red and black lines provide the transmission of the neutral gas and the entire ISM, respectively
               (see also Sect.~\ref{sec:GX3394}).}}
          \label{fig:GX339_transmission}
    \end{figure}

{In this section we show the detailed analysis of the two GX~339-4 stacked spectra between 15 and 24\,{\AA}. This will also indicate the relevance of each absorber in the final model. We choose to show this source because of its highest statistics and large column density that maximize the significance of the spectral features. Moreover, the results obtained for GX~339-4 are similar to those of most sources in our sample. We first jointly fit the Fe L and O K edges ($15-24$\,{\AA}) with a simple model consisting of a power-law absorbed by neutral gas. Free parameters are the normalization and slope of the power-law and the Fe and O abundances of the neutral gas. This fit provides $\chi^2_{\nu}=1692/454=3.73$, which is clearly poor and unacceptable (see also the blue line in Fig.~\ref{fig:GX339_edges}). A different continuum does not produce any change. Therefore, we add ions in sequence from \ion{O}{ii} to \ion{O}{viii}, see Table~\ref{table:GX339}. This (neutral + ionized) gas model with all these 
ions gives $\chi^2_{\nu}=1092/447=2.44$ and a big improvement to the fit. However, the fit is not yet acceptable and there are residuals at 17.5 and 22.9\,{\AA} (see red line). Moreover, the pure gas model does not reproduce the ratio between the absorption lines and edges for both iron and oxygen. Dust is definitely needed. We complete the ISM model by adding in sequence different molecules with the order as suggested by preliminary fits to each edge, which is mentioned above: metallic iron, water ice, pyroxene silicates, up to iron sulfate (see also Table~\ref{table:GX339}). The final $\chi^2_{\nu}$ drops down to about 1.7 as shown by the final model (purple line in Fig.~\ref{fig:GX339_edges}). Although this fit is not yet formally acceptable, the final $\chi^2_{\nu}$ (significantly higher than one) is mainly due to the high statistics. Similar $\chi^2_{\nu}$ values are obtained for featureless regions at similar statistical quality and indicate the final systematic uncertainties of the instrument model. 
In Fig.~\ref{fig:GX339_transmission} we show the transmission of the ISM best-fitting model near the \ion{O}{i} edge of GX~339-4. We also report the relative contributions provided by \ion{O}{i} (red solid line), \ion{O}{ii} (purple dotted line), \ion{O}{iii-iv} (black dashed line), as well as by specific compounds. Molecular compounds dominate the spectral range within 22.8 -- 23.2\,{\AA} and affect the ratio between the depths of the \ion{O}{i} edge and resonance line.}

\begin{table}
\caption{GX 339-4 spectral fits within $15-24$\,{\AA} and relative $\Delta\,\chi^2$.}
\renewcommand{\arraystretch}{1.3}
\label{table:GX339}
%  \small\addtolength{\tabcolsep}{-1pt}
\begin{tabular}{lr|rr|rr}
\hline
Parameter     & $\Delta\,\chi^2$ & Parameter  & $\Delta\,\chi^2$ & Parameter & $\Delta\,\chi^2$  \\
\hline
\ion{O}{ii}   &  14    & Metallic Fe                      & 101         & CO            & 1 \\
\ion{O}{iii}  & 163    & H$_2$\,O ice                     & 168         & Fe\,O\,(OH)   & 1 \\
\ion{O}{iv}   &  20    & Mg\,Si\,O$_3$                    &  4$^{(a)}$  & N$_2$O        & 0 \\
\ion{O}{v}    &   1    & Fe$_3$\,O$_4$                    &  60         & Fe$_2$\,O$_3$ & 0 \\
\ion{O}{vi}   &   1    & Ca$_3$\,Fe$_2$\,Si$_3$\,O$_{12}$ &   3         & Fe\,SO$_4$    & 0 \\
\ion{O}{vii}  & 366    & Fe\,Al$_2$\,O$_4$                &   2         &               &  \\
\ion{O}{viii} &  35    & Mg$_{1.6}$\,Fe$_{0.4}$\,SiO$_4$  &   2         &               &  
\end{tabular}
\vspace{0.1cm}

          {The $\Delta\,\chi^2$ values refer to the sequential changes obtained by adding the
           parameters from top to bottom and from left to right: \ion{O}{ii}, \ion{O}{iii}, ..., \ion{O}{viii}, Metallic Fe, ..., Fe\,SO$_4$.
           $^{(a)}$ The $\Delta\,\chi^2$ given by pyroxene Mg\,Si\,O$_3$ is small because its features are similar to 
           those produced by H$_2$\,O ice (see also Fig.~\ref{fig:GX339_contours}).}
\end{table}

{Once we obtain a good fit for the $15-24$\,{\AA} spectral range, we fit the model to the entire spectrum. Additional free parameters in the model are the neon and magnesium abundances of the cold gas ({\sl hot} model in SPEX) and the column densities of \ion{Ne}{ii-iii}, \ion{Ne}{viii-x}, \ion{Mg}{xi-xii}, and \ion{Fe}{xvii-xviii} (which are provided by the {\sl slab} model).}

\subsection{{Results}}
\label{sec:results}

{This procedure used for GX~339-4 is identically applied to the other eight targets and} the results of our best-fitting models obtained on all the X-ray sources are shown in Table~\ref{table:fit_complete} and plotted in Figs.~\ref{fig:neon}--\ref{fig:oxygen}. The model fits very well both the absorption edges and lines. {The reduced $\chi^2$ for the nine sources are between 1.4 and 1.7}. We show the absolute abundances and hydrogen column density for the cold gas, the column densities for all ions and molecules that we were able to detect. In the cases of no detection we report the 2\,$\sigma$ upper limits. {We also report the formula for each compound and the total amount of iron and oxygen found in dust}. We will discuss the results in Sect.~\ref{sec:discussion}. Briefly, we have found a general agreement in the spectra of the different sources. Ions like \ion{O}{v-vi} and molecular compounds like hematite (Fe$_2$O$_3$), and CO are hard to detect. Instead, \ion{O}{vii-viii}, H$_2$O ices, and compounds of Ca (andradite), and Al (Hercinite) are detected along any LOS. The cold neutral gas provides on average \ion{O}{i} $\sim1-3\times10^{22}$\,m$^{-2}$, which means that most oxygen is found in the neutral phase. We have summed all the contributions to the molecular oxygen as well as all the ionic column densities of the mildly ionized (\ion{O}{ii-v}) and heavily ionized (\ion{O}{vi-viii}) gas, and compare them in Table~\ref{table:fit_colfrac}. As we have mentioned above, most of the oxygen is provided by the neutral gas phase. Dust also seems to be ubiquitous. The warm and hot ionized gas phases generally contribute less to oxygen, but there are some cases in which they constitute significant portions of the 
total column density. {SAX~J1808.4--3658 is an exceptional case with detectable \ion{O}{vi} and prominent \ion{O}{vii} absorption lines (see also Fig.~\ref{fig:oxygen}, bottom-right panel).}
 
\begin{landscape} %LANDSCAPE BEGIN
\begin{table}
\renewcommand{\arraystretch}{1.3}
\caption{RGS fits with the complete ISM model. {The dust and the three gas phases are represented by the four main blocks.}}
\begin{center}
% use packages: array
 \small\addtolength{\tabcolsep}{-1pt}
\scalebox{0.95}{%
\begin{tabular}{llllllllll}
 \hline
Parameter                  & 4U1254--690      & 4U 1636--536     & 4U 1735--444    & Aql X--1        & GS 1826--238     & GX 339--4        &  GX 9+9          & SAX J1808.4--3658         &   Ser X-1        \\                                    
\hline
 $N_{\rm H}\,(10^{\,25}$ m$^{-2})$ 
                         & 2.69 $\pm$ 0.03 & 3.58 $\pm$ 0.07   & 3.28 $\pm$ 0.08 & 5.21  $\pm$ 0.05  & $4.14\pm0.06$   & 5.1   $\pm$ 0.4   & 2.15 $\pm$ 0.05 & 1.40   $\pm$ 0.03 &  5.0 $\pm$ 0.3 \\
 O/H$\,^{(a)}$           & 0.92 $\pm$ 0.04 & $0.85\pm0.10$     & 0.93 $\pm$ 0.03 & 0.88  $\pm$ 0.02  & $0.92 \pm0.04$  & 1.06  $\pm$ 0.05  & 0.99 $\pm$ 0.03 & $0.91 \pm 0.03$   &  0.9 $\pm$ 0.1 \\
 Ne/H$\,^{(a)}$          & 0.91 $\pm$ 0.03 & $1.31\pm0.05$     & 1.00 $\pm$ 0.05 & 0.95  $\pm$ 0.07  & $1.23\pm0.05$   & 1.05  $\pm$ 0.05  & 1.35 $\pm$ 0.07 & $1.37 \pm 0.05$   &  $1.20\pm0.05$ \\
 Mg/H$\,^{(a)}$          & 1.7 $\pm\,_{1.0}^{0.3}$ &   $<$ 0.8 & 2.0 $\pm\,_{1.0}^{0.3}$ & 1.0 $\pm\,_{0.8}^{0.2}$ & 1.4 $\pm\,_{0.6}^{0.2}$ &$1.2\,\pm\,_{0.8}^{0.3}$& 0.4 $\pm\,_{0.3}^{0.5}$ & $<$ 0.2 & 0.9 $\pm\,_{0.1}^{0.5}$\\
 Fe/H$\,^{(a)}$ & 0.16 $\pm\,_{0.1}^{0.05}$& 0.18 $\pm$ 0.04   & 0.25 $\pm$ 0.10 & 0.11  $\pm$ 0.05  & $0.22\pm0.06$   & 0.4   $\pm$ 0.1   & 0.19 $\pm$ 0.05 & $<$ 0.1           &  $0.50 \pm 0.05$	 \\
\hline
 \ion{O}{ii}$\,^{(b)}$     &   0.33 $\pm$ 0.15 &   1.4 $\pm$ 0.7 & 0.6 $\pm$ 0.2   &  0.2 $\pm$ 0.1  &  0.6  $\pm$ 0.1  &  1.4 $\pm$ 0.5   & 0.3  $\pm$ 0.1  & 1.67 $\pm$ 0.05 & 2.0  $\pm$ 0.2 \\ 
 \ion{O}{iii}$\,^{(b)}$    & 0.08 $\pm\,_{0.04}^{0.08}$ & $<$ 0.1& 0.2 $\pm$ 0.1   &    $<$ 0.3      &    $<$ 0.3       &     $<$ 0.8      &     $<$ 0.05    & 0.34 $\pm$ 0.05 & $<$ 0.4        \\ 
 \ion{O}{iv}$\,^{(b)}$     &   0.5 $\pm$ 0.2   & 0.3 $\pm$ 0.2   & 0.4 $\pm$ 0.2   &  0.4  $\pm$ 0.1 &  0.6  $\pm$ 0.2  &  0.5 $\pm$ 0.1   & 0.4  $\pm$ 0.2  & 0.31 $\pm$ 0.03 & 0.7  $\pm$ 0.5  \\
 \ion{O}{v}$\,^{(b)}$      &     $<$ 0.04      &    $<$ 0.02     &  $<$ 0.1        &   $<$ 0.03      &   $<$ 0.06       &      $<$ 0.01    &     $<$ 0.002   &0.006 $\pm$ 0.002& 0.3  $\pm$ 0.2  \\
 \ion{Ne}{ii}$\,^{(b)}$    &  0.11 $\pm$ 0.06  & 1.2  $\pm$ 0.3  & 0.4  $\pm$ 0.2  &  0.7 $\pm$ 0.2  &  1.4 $\pm$ 0.3   &  1.2 $\pm$ 0.5   & 0.4 $\pm$ 0.2   & 1.3  $\pm$ 0.1  & 0.3  $\pm$ 0.2 \\  
 \ion{Ne}{iii}$\,^{(b)}$   &  0.19  $\pm$ 0.08 & 0.9  $\pm$ 0.2  & 0.5  $\pm$ 0.2  &  0.6 $\pm$ 0.2  &  0.6 $\pm$ 0.3   &  0.9 $\pm$ 0.2   & 0.2 $\pm$ 0.1   & 1.1  $\pm$ 0.1  & 0.8  $\pm$ 0.4  \\  
\hline
 \ion{O}{vi}$\,^{(b)}$     &     $<$ 0.03      &    $<$ 0.06     &  $<$ 0.01       &   $<$ 0.06      &   $<$ 0.07       &      $<$ 0.04    &     $<$ 0.004   & 0.02 $\pm$ 0.01 &      $<$ 0.05  \\
 \ion{O}{vii}$\,^{(b)}$    & 0.10  $\pm$ 0.05  & 0.20 $\pm$ 0.07 & 0.2 $\pm$ 0.1   & 0.15 $\pm$ 0.06 &  0.15 $\pm$ 0.06 &  0.4 $\pm$ 0.1   & 0.04 $\pm$ 0.02 & 2.2  $\pm$ 0.5  & 0.24 $\pm$ 0.05 \\
 \ion{O}{viii}$\,^{(b)}$   & 0.09 $\pm$ 0.04   & 0.19 $\pm$ 0.05 & 0.3 $\pm$ 0.1   & 0.10 $\pm$ 0.04 &  0.10 $\pm$ 0.03 &  0.3 $\pm$ 0.1   &     $<$ 0.005   & 0.11 $\pm$ 0.02 & 0.20 $\pm$ 0.06 \\   
 \ion{Ne}{viii}$\,^{(b)}$  &       $<$ 0.03    & 0.07 $\pm$ 0.04 &      $<$ 0.05   &      $<$ 0.01   &    $<$ 0.01      &  0.06 $\pm$ 0.04 &    $<$ 0.004    &      $<$ 0.003  &       $<$ 0.03 \\
 \ion{Ne}{ix}$\,^{(b)}$    & 0.05 $\pm$ 0.03   & 0.09 $\pm$ 0.04 &  0.2 $\pm$ 0.1  & 0.07 $\pm$ 0.05 & 0.06 $\pm$ 0.02  &  0.16 $\pm$ 0.04 & 0.03 $\pm$ 0.02 & 0.05 $\pm$ 0.02 & 0.14 $\pm$ 0.05 \\
 \ion{Ne}{x}$\,^{(b)}$     &  $<$ 0.04         &      $<$ 0.03   &  $<$ 0.02       &  $<$ 0.1        &    $<$ 0.02      &  $<$ 0.02        &      $<$ 0.006  &      $<$ 0.008  &      $<$ 0.04  \\
%\hline
 \ion{Mg}{xi}$\,^{(b)}$    &  0.16 $\pm$ 0.05  & 0.08 $\pm$ 0.04 & 0.4 $\pm$ 0.3   &  $<$ 0.05       &    $<$ 0.1       &  0.10 $\pm$ 0.05 & 0.06 $\pm$ 0.03 & 0.18 $\pm$ 0.09 &       $<$ 0.15 \\
 \ion{Mg}{xii}$\,^{(b)}$   &    $<$ 0.01       &   $<$ 0.01      &  $<$ 0.01       &  $<$ 0.02       &    $<$ 0.02      &    $<$ 0.01      &     $<$ 0.005   &      $<$ 0.01   &       $<$ 0.03 \\
 \ion{Fe}{xvii}$\,^{(b)}$  &    $<$ 0.01       & 0.02 $\pm$ 0.01 & 0.02 $\pm$ 0.01 & 0.02 $\pm$ 0.01 & 0.02 $\pm$ 0.01  & 0.02 $\pm$ 0.01  &0.008 $\pm$ 0.004& 0.02 $\pm$ 0.01 &       $<$ 0.01 \\
 \ion{Fe}{xviii}$\,^{(b)}$ &    $<$ 0.02       &    $<$ 0.01     & $<$ 0.01        & 0.02 $\pm$ 0.01 &  $<$ 0.01        &    $<$ 0.01      &      $<$ 0.003  &      $<$ 0.04   &       $<$ 0.01 \\
\hline
 Water ice H$_2$\,O $\,^{(b)}$                            & 2 $\pm\,{1}$      & 0.9  $\pm$  0.5   &  3  $\pm$  2      &  3    $\pm$  1    &    3  $\pm$ 2     &  4.4  $\pm$ 2.5   &   2  $\pm$  1     & 0.3 $\pm\,_{0.1}^{0.7}$& $<{6.6}$ \\
 Andradite Ca$_3$\,Fe$_2$\,Si$_3$\,O$_{12}$ $\,^{(b)}$    & 0.02  $\pm$ 0.01  & 0.02 $\pm$ 0.01   & 0.02 $\pm$ 0.01   & 0.03  $\pm$ 0.01  & 0.03  $\pm$ 0.01  & 0.03  $\pm$ 0.01  & 0.012 $\pm$ 0.004 & 0.008$\pm$ 0.004& 0.03 $\pm$ 0.01 \\
 Hercinite Fe\,Al$_2$\,O$_4$ $\,^{(b)}$                   & 0.04  $\pm$ 0.02  & 0.05  $\pm$  0.02 & 0.04  $\pm$ 0.02  &        $<$  0.025 & 0.06  $\pm$ 0.02  & 0.07  $\pm$ 0.02  & 0.03  $\pm$ 0.01  & 0.015$\pm$ 0.007& 0.07 $\pm$ 0.02 \\
 Lepidocrocite Fe\,O\,(OH) $\,^{(b)}$                     & 0.11  $\pm$  0.06 & 0.4   $\pm$  0.2  &   $<$ 0.2         &  0.6  $\pm$ 0.3   &  0.4  $\pm$ 0.2   & 0.03  $\pm$ 0.02  &  0.2 $\pm$ 0.1    & 0.15 $\pm$ 0.08 &  0.15 $\pm$ 0.10\\
 Pyroxene Mg\,Si\,O$_3$ $\,^{(b)}$                        &  $<$ 1.2          &  1.2 $\pm$ 0.5    &   $<$ 1.1         &   $<$ 1.2         &  $<$ 1.1          & 0.3 $\pm\,_{0.1}^{0.7}$ & $<$ 1.0     & $<$ 0.4         &  1.7  $\pm$ 0.5 \\
 Carbon monoxide CO $\,^{(b)}$                            &  $<$ 0.11         &  0.3 $\pm$ 0.2    &   $<$ 0.2         &   $<$ 0.1         &  $<$ 0.3          &  0.07 $\pm$ 0.05  &   $<$ 0.05        & 0.6  $\pm$ 0.1  &   $<$ 0.2       \\
 Metallic Fe $\,^{(b)}$                             &0.3 $\pm\,_{0.1}^{0.3}$&0.3 $\pm\,_{0.1}^{0.3}$&0.3 $\pm\,_{0.1}^{0.2}$&$<$ 0.7      &  $<$ 1.2          &  0.6 $\pm$ 0.1    & 0.3 $\pm\,_{0.1}^{0.3}$& 0.13 $\pm$ 0.05&   $<$ 0.7   \\
 Magnetite Fe$_3$\,O$_4$ $\,^{(b)}$                       & 0.03 $\pm$ 0.02   &  0.2 $\pm$ 0.1    &   $<$ 0.04        &  0.3  $\pm$ 0.2   &  0.3  $\pm$ 0.1   &  0.2  $\pm$ 0.1   &   $<$ 0.02        & $<$ 0.005       & 0.2  $\pm$ 0.1  \\
 Olivine Mg$_{1.6}$\,Fe$_{0.4}$\,SiO$_4$ $\,^{(b)}$       &  $<$ 0.04         &  0.06 $\pm$ 0.03  &   $<$ 0.1         &   $<$ 0.04        &  $<$ 0.08         &  0.07 $\pm$ 0.03  &   0.2 $\pm$ 0.1   & $<$ 0.03        &   $<$ 0.1       \\
 Laughing gas N$_2$O $\,^{(b)}$                           &  0.4  $\pm$ 0.3   &       $<$ 0.04    &   $<$ 0.1         &   $<$ 0.03        &  $<$ 0.2          &  $<$ 0.03         &   $<$ 0.03        & 0.03 $\pm$ 0.02 &   $<$ 0.04      \\
 Hematite Fe$_2$\,O$_3$ $\,^{(b)}$                        &  $<$ 0.04         &       $<$ 0.01    &   $<$ 0.1         &   $<$ 0.02        &  $<$ 0.2          &  $<$ 0.003        &   $<$ 0.02        & $<$ 0.01        &   $<$ 0.02      \\
 Iron sulfate Fe\,SO$_4$ $\,^{(b)}$                       &  $<$ 0.04         &       $<$ 0.01    &   $<$ 0.05        &   $<$ 0.03        &  $<$ 0.01         &  $<$ 0.004        &   $<$ 0.04        & $<$ 0.03        &   $<$ 0.02      \\
\hline                                                                                                            
 Total oxygen in dust $\,^{(b)}$ & 3 $\pm$ 1     & 7 $\pm$ 2     & 4 $\pm$ 1     & 6 $\pm$     1 & 6 $\pm$ 1     & 7 $\pm$ 2     & 4 $\pm$ 1     & 1.5 $\pm$ 0.5 & 7 $\pm$ 2 \\
 Total iron in dust $\,^{(b)}$   & 0.6 $\pm$ 0.2 & 1.3 $\pm$ 0.5 & 0.5 $\pm$ 0.3 & 1.7 $\pm$ 0.5 & 1.5 $\pm$ 0.6 & 1.4 $\pm$ 0.5 & 0.7 $\pm$ 0.2 & 0.3 $\pm$ 0.1 & 0.8 $\pm$ 0.4 \\
\hline                                                                                                            
\end{tabular}}
\label{table:fit_complete}
\end{center}
$^{(a)}$ Abundances ratios are in the linear proto-Solar abundance units of \citet{Lodders09}.\\
$^{(b)}$ Ionic and molecular column densities are in units of $10^{\,21}$\,m$^{-2}$ (see also Sect.~\ref{sec:analysis}).\\
\end{table}
\end{landscape} %LANDSCAPE END

\subsection{{Limitations}}
\label{sec:limitations}

\begin{figure}
      \centering
      \includegraphics[width=6.6cm, bb=65 85 525 700, angle=+90]{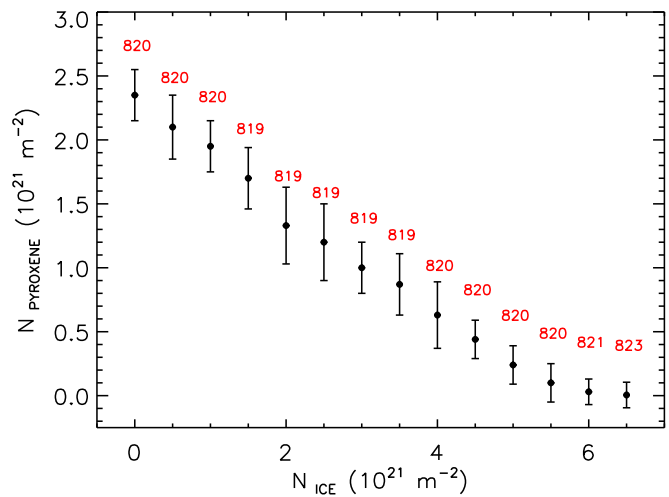}%}
      \caption{{Water ice / pyroxene anti-correlation. The pyroxene column densities refer
                                 to the spectral fits of the Fe L and O K edges for GX~339-4 obtained with fixed amount of water ice. 
                                 The red values show the $\chi^2$ values of each fit.
                                 For more detail see Sect.~\ref{sec:GX3394} and \ref{sec:limitations}.}}
          \label{fig:GX339_contours}
    \end{figure}

{There are some factors that may limit and affect the uniqueness of our best-fitting models. First of all, the features produced by molecules in a certain absorption edge are similar and degenerate. {Near the Fe L-edge the RGS absorption lines blend and may not give unique solutions. \textit{Chandra} spectra are provided with higher resolution, but they give accurate results only for a few sources even brighter than ours.} Therefore, in our standard modeling we have simultaneously fitted all the molecules and all the absorption edges and lines in order to obtain the best-fitting mixture and to break as much as possible the degeneracy. Unfortunately, for some compounds, such as magnetite and pyroxene, we only have the transmission near the oxygen edge, while for the other edges we simply use the pure atomic cross-section without absorption lines (for details consult the {\sl amol} model in the SPEX manual). {Fortunately, these systematic effects do not strongly alter the depletion factors, especially 
for oxygen. The Mg K-edge is very weak and it does not make any difference whether we model it with atomic or molecular cross-sections. Another problem is the degeneracy between the absorption features produced by water ice and pyroxene (see Fig.~\ref{fig:GX339_transmission} and \ref{fig:GX339_contours} and \citealt{Costantini2012},  Appendix~A), which are the molecules that best fit the O K-edge. This is also shown by the large uncertainties on their column densities in Table~\ref{table:fit_complete}. Therefore, in order to test the robustness of our results and to estimate the systematic effects, we have performed additional fits with different combinations of molecules by excluding some of them. One model includes only the molecules that do not lack any relevant cross-section, such as water ice, Metallic Fe, CO, and Hematite (all the other molecules such as pyroxene and magnetite are excluded). In another test we replace water ice with pyroxene in order to measure systematic changes in the Mg abundance of 
the cold gas. Several tests like these provided us with systematic changes in the column densities of the molecules and the ions as well as in the abundances of the cold gas. We have summed all the systematic and statistic errors and reported them together with the best-fitting values in Table~\ref{table:fit_complete}.} Interestingly, the total amount of oxygen and iron in dust does not strongly depend on the chosen dust mixture. Therefore, their depletion factors are not highly effected, but the detailed chemical structure is.}

On another hand, the molecular database is not complete. We have tested all the molecules available in the SPEX database, which are about 40 (carbon oxides, ices, iron and magnesium silicates, and more complex molecules). Although these compounds are among the most abundant in the ISM, we could still miss important contribution from other species like forsterite (Mg$_2$\,Si\,O$_4$, see e.g. \citealt{Jones2000}). {Our ISM team at SRON is currently taking synchrotron and electroscopic measurements of the cross-sections near the O, Mg, and Si K-edges, and both the Fe K\,/\,L edges of several silicate compounds with a high spectral resolution too. These laboratory data will provide accurate estimates of molecular column densities.}

{The results obtained on the various absorption edges have different weights. The oxygen edge is the deepest and the column densities of oxygen compounds are constrained better than for iron. The Mg edge is very weak for hydrogen column densities like those of our sources and the estimates of magnesium abundances and depletion factors are much more uncertain. This does not apply to neon as it is supposed to be only in a gaseous phase. This work mostly focuses on measurements of interstellar oxygen, iron, and neon column densities. A deep analysis of magnesium and silicon would require the use of both the \textit{XMM-Newton} RGS and the \textit{Chandra} HETG gratings (which have higher effective area and resolution than RGS at short wavelength) in combination with sources with higher hydrogen column densities.}

\begin{figure}
%     \centering
%  \vspace{-0.5cm}
\includegraphics[width=0.35\textwidth, angle=90, bb=55 70 560 740]{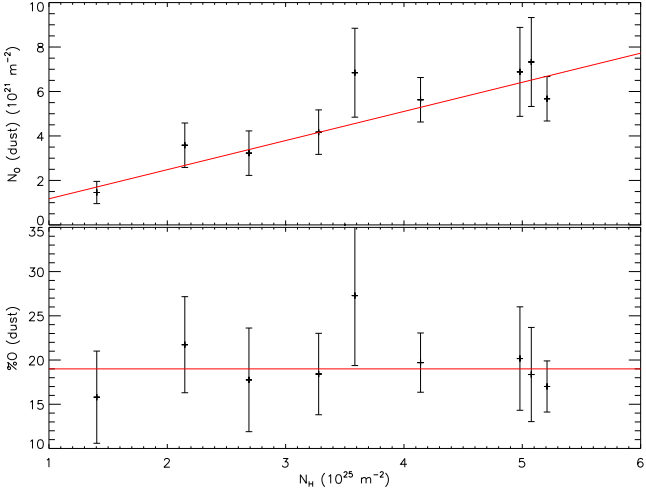}
\caption{Amount of \ion{O}{i} in dust and its fraction with respect to the total neutral oxygen (gas + dust) versus the hydrogen column density.}
    \label{fig:oxygen_amol}
   \end{figure}

\begin{figure}
    \centering
%  \vspace{-0.5cm}
\includegraphics[width=0.34\textwidth, angle=90, bb=55 55 560 740]{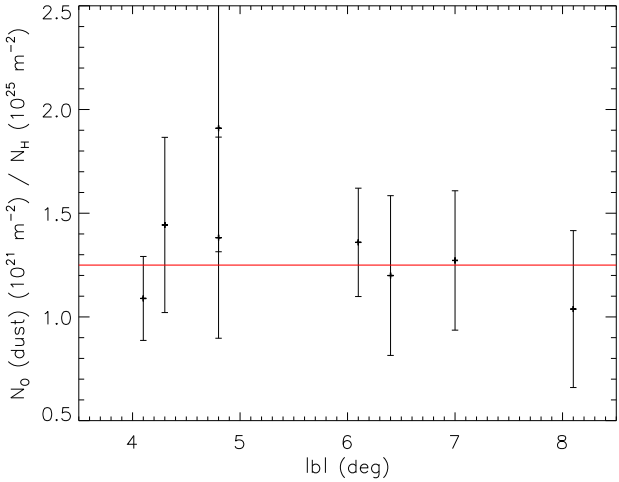}
\caption{Ratio between the Amount of \ion{O}{i} in dust and the hydrogen column density versus the absolute value of the Galactic latitude.}
    \label{fig:oxygen_amol_gas_ratio}
   \end{figure}

{Our estimates of column densities and abundances provide interstellar properties as integrated along the LOS and on a large scale. This is reasonable for the diffuse gas, which is almost uniformly distributed between 3--15\,kpc from the Galactic Center and whose warm phase may reach a few kpc height from the Galactic place \cite[see e.g.][]{ferriere}. This scale-height is indeed comparable to the average altitude of our sources (see Table~\ref{table:1}). However, this assumption may not be correct for dust and molecules, which follow the Galactic spirals and are usually found at lower altitudes. This means that dust sampling towards our targets may be closer to the Sun rather than in the middle of the path. In the cold phase, iron and magnesium are strongly depleted into dust. Therefore, the average location of the Fe and Mg absorbers can be found at shorter distances than those referring to oxygen and neon. {However, in Sect.~\ref{sec:ISM_structure} we will show that along our low-latitude LOS the 
different scale-heights of gas and dust will not affect the results.}

{Another problem in our grating spectra is the presence of several bad pixels or dead columns that affect the estimates of some parameters. One of the most relevant is located at $22.75$\,{\AA}, just near the \ion{O}{iv} resonance line (see Figs.~\ref{fig:oxygen} and \ref{fig:GX339_transmission}). The \ion{O}{iv} column densities of our targets are all determined just through this line and their values may be inaccurate. New observations taken with the multi-pointing RGS mode will surely provide a solution to this problem.}

\section{Discussion}
\label{sec:discussion}

Our analysis shows that the ISM has a multi-phase structure characterized by cold neutral gas (and dust), mildly ionized gas and heavily ionized gas. This complex structure is found towards any of the studied sources (see Table~\ref{table:fit_complete}) and is consistent with our previous work \citep[see e.g.][]{kaastra09, Pinto2010, Pinto2012, Costantini2012}. Signatures of dust are always present with main contributions by ices and silicates. The oxygen edge and lines provide the most accurate results due to their large depth, but the results are even more robust because of the simultaneous modeling of all the absorption edges. This was crucial especially for fitting dust and molecules which affect both O, Fe, and Mg edges. An interesting and simple check of different amounts of oxygen in dust is provided by the comparison of the O\,K edge in Fig.~\ref{fig:oxygen}. As we anticipated in Sect.~\ref{sec:spectra}, the \ion{O}{i} 1s--2p absorption line is deeper than the oxygen edge for SAX~J1808.4$-$3658, 
while for Ser~X--1 and GX~339--4 they are comparable. We attribute this to oxygen depleted from the gas phase into dust grains, which is confirmed by the measurement of a larger amount of oxygen in solids in the LOS towards the two latter sources (see Table~\ref{table:fit_complete}).

\subsection{The nature of the gas phases}
\label{sec:alternative_models}

In our standard model we have used a collisionally-ionized model in SPEX because at low temperatures this mimics well the cold interstellar gas. However, at these temperatures photo-ionization could also provide a significant contribution. Therefore, we tried an alternative model in which a photo-ionized component ({\sl xabs} in SPEX) substitutes the cold collisionally-ionized {\sl hot} component. The fits are equivalent, the gas is almost neutral with a very low ionization parameter $\xi$. Abundances and $\chi^2$ are consistent with our standard model. The main difference is that the fit with the photo-ionized model takes more time to converge and the absolute abundances, i.e. relative to hydrogen, have larger uncertainties \citep[see also][]{Pinto2012}. Therefore, we prefer to keep our standard model even if photo-ionization is not ruled out.

We have also considered an alternative, physical model for the warm mildly-ionized gas that produces most of the \ion{O}{ii-iv} and \ion{Ne}{ii-iii} absorption lines. Instead of a {\sl slab} absorber we have tested a photo-ionized {\sl xabs} component, which was successfully used by \citet{Pinto2012} in the deep UV and X-ray spectra of the AGN Mrk~509. The fit was poor because a single photo-ionized component is not able to fit all the large column densities as measured by our empirical standard model and shown in Table~\ref{table:fit_complete}. We need a low-ionization component which provides the bulk of \ion{O}{ii} and \ion{Ne}{ii} together with an intermediate-ionization component for the other ions. Interestingly, a good fit is obtained when using the same abundance pattern as provided by the cold neutral gas, while proto-Solar abundances give a bad fit. However, a thorough study of the warm gas requires longer exposures and complementary UV data, which is beyond the scope of this paper.
 
The heavily ionized gas is usually found to be in collisional equilibrium. However, it is thought that this hot gas in the Galaxy is not characterized by a single phase, but it is constituted of two main phases. Most \ion{O}{vii-viii} should arise from a very hot ($\sim2\times10^6$\,K) collisionally-ionized phase, while the bulk of \ion{O}{vi} is thought to be embedded in a cooling phase with temperatures $1-5\times 10^5$\,K \citep[see e.g.][]{Richter2006}. We have tested both single-phase and two-phase models by substituting the {\sl slab} component responsible for the \ion{O}{vi-viii} absorption with one and two collisionally-ionized {\sl hot} components in SPEX. The fits are comparable due to the weakness of their absorption lines. We would need UV \ion{O}{vi} data in order to put strong constraints, but most of our targets have not been observed with COS/HST or FUSE. However, Table~\ref{table:fit_complete} shows a spread larger than 10 in the \ion{O}{vii}~/~\ion{O}{viii} column ratio, which for a single 
gas component would provide a very large scatter in temperature between the nine LOS (from 1.2 to 2.6 million K). We think that combinations of two stable, but different hot (cooling and coronal) phases is a more likely description.

\subsection{Is the ISM chemically homogeneous ?}
\label{sec:ISM_structure}

On small scales the ISM is not expected to be highly homogeneous because there are significant physical and chemical differences between the interstellar environments like \ion{H}{ii} regions, PNe nebulae, dark clouds, etc. However, the diffuse medium in our Galaxy may be homogeneous on large scales. Some chemical and physical properties may be possibly consistent when comparing distant regions which underwent a similar evolution. This is possible for dust as most of it is expected to grow in the ISM rather than in stars \citep[see e.g.][]{Mattsson2012}. It is thus worth to compare its chemical properties as integrated along the different LOS. First of all, we notice that the molecular composition is similar in any direction. Aluminates and calcium silicates are found to provide most of Al and Ca, which are indeed expected to be highly depleted into dust. {Water ice may be present, but in an uncertain amount due to the degeneracy between its features and those of pyroxene. \citet{Jenkins2009} 
suggested water ice as a possible principal resource of oxygen in dust, but he also showed that at low-intermediate $N_{\rm H}$ it is hard to detect.} Cross-sections of these and of additional compounds are currently being measured with a resolution higher than the current one and may provide more accurate results (see Sect.~\ref{sec:limitations}).} To understand the role of dust, we have summed all the contributions to oxygen from solids and molecules and compared it with the hydrogen column density that we have obtained in our best fits (see Fig.~\ref{fig:oxygen_amol}). We show the trends of both the total column density of \textit{dusty} oxygen and its fraction with respect to the total amount of neutral oxygen (gas + dust) as function of the N$_{\rm H}$. The oxygen dust column correlates well with the hydrogen column density (slope $=(1.3\pm0.2) \times 10^{-4}$) and the fraction of oxygen bound in dust is constant ($19\pm5$\%) along any LOS. These results suggest that on average the cold ISM phase is 
chemically homogeneous. We notice that the total amount of oxygen in dust is solidly measured as it does not strongly depend on the dust mixture. For instance, we tested different molecular combinations by removing some of the main compounds from the fit, but in the end the sum of all the oxygen molecular column densities provide the same results in the fits. The same applies for iron, whose dust fraction does not depend on the N$_{\rm H}$ and the LOS, but it has a larger uncertainty and spreads between 65--90\%. This occurs because the iron L edge is shallower than the oxygen K edge. 
{As we mentioned in Sect.~\ref{sec:limitations}, the scale-height of the interstellar dust is smaller than for the gas, thus, due to the altitudes of the targets, the dust is expected at shorter distances from the Sun. This means that the analysis of the dust in these different LOS probes properties of the ISM as measured on a large scale, but probably smaller than with the gas. {As additional check we have computed the dust-to-gas ratio for our sources as the ratio between the total amount of oxygen in dust and the hydrogen column density. The dust-to-gas ratio is constant (slope $=0.0\pm0.2 \times 10^{-4}$) at any value of Galactic latitude as shown in Fig.~\ref{fig:oxygen_amol_gas_ratio}. This result and the trend seen in Fig.~\ref{fig:oxygen_amol} suggest that the distribution of the dust should not deviate from the gas. This may be due to the low latitudes of our sources, which was also one of the reasons for us to choose these targets as representatives of the Galactic (inner) disk.}}

 \subsection{ISM column densities}
 \label{sec:discuss_column}

Our empirical model provides us with estimates of column densities for several molecular and ionic species (see Table~\ref{table:fit_complete}). In some cases we could only obtain upper limits, but we prefer to show them as they may be useful for future work. In order to further check the robustness of our model we compare our column density estimates with those found in the literature. In most cases our best-fit values of $N_{\rm H}$ agree with those measured at longer wavelength (see Table~\ref{table:1}). On average our values are slightly higher that the 21\,cm \ion{H}{i} estimates probably because the curvature as measured in the X-ray above 25\,{\AA} takes into account the contributions from \ion{H}{i} and H$_2$. Our values are consistent with other papers \citep[see e.g.][]{Xiang2005, Caballero2009, Bhattacharyya2007, Patruno2009}. Our column density estimate for GS\,1826--238 agrees with what we found with a different model in \citet{Pinto2010}, i.e. $N_{\rm H} = 4.14 \pm 0.07 \times 10^{25}$\,m$^{-2}
$. All the neutral column densities that we have measured for 4U~1636--536, 4U~1735-444, GX~339--4, GX~9+9, and Ser~X-1 are consistent with \citet{JuettI, JuettII}. We found a disagreement between our average values of \ion{O}{ii}/\ion{O}{i} and \ion{O}{iii}/\ion{O}{i} column density ratios, $\sim$0.06 and $\sim$0.02 respectively, with those obtained by \citet{JuettI} of about 0.1. This may be due to the fact that their model does not take into account any contribution by dust or molecules near 23\,{\AA} and thus may overestimate the amount of ionized oxygen. {They obtained acceptable fits without including dust in their models, but this was due to their short exposures and to the sensitivity of \textit{Chandra} which is smaller than \textit{XMM-Newton} near the O K-edge.
Our ionic column densities agree very well with those estimated by \citet{Miller2004} for GX~339--4. According to our best fit models, \ion{O}{iv} appears to have column densities larger than \ion{O}{iii}. The most obvious cause may be the presence of a bad pixel exactly near the \ion{O}{iv} resonance line as mentioned in Sect.~\ref{sec:limitations}. However, it may also be that we have slightly over-estimated the absorption of molecules and dust near at 21\,{\AA} at expensive of the \ion{O}{iii} amount. Recently, the \textit{XMM-Newton}/RGS has been provided with an optimal observation scheme, the multi-pointing mode, which allows snapshots of the same source with shifts in the dispersion direction. These shifts place the bad pixels in different location and allows to reconstruct a completely clean spectrum. This procedure has been successfully used in large campaign on AGN Mrk~509 (see e.g. \citealt{Kaastra11b}) and provided our team with a bright and clean spectrum. Once applied to sources with high $N_{\
rm H}$ like LMXB, this procedure will reveal the composition of the ISM in detail.} We point out that many of the high-resolution X-ray (RGS) spectra of these sources have been analyzed for the first time by us.

\subsection{ISM abundances}
\label{section:ISM_abundances}

\begin{figure*}
    \centering
%  \vspace{-0.5cm}
\includegraphics[width=0.6\textwidth, angle=90, bb=55 55 540 720]{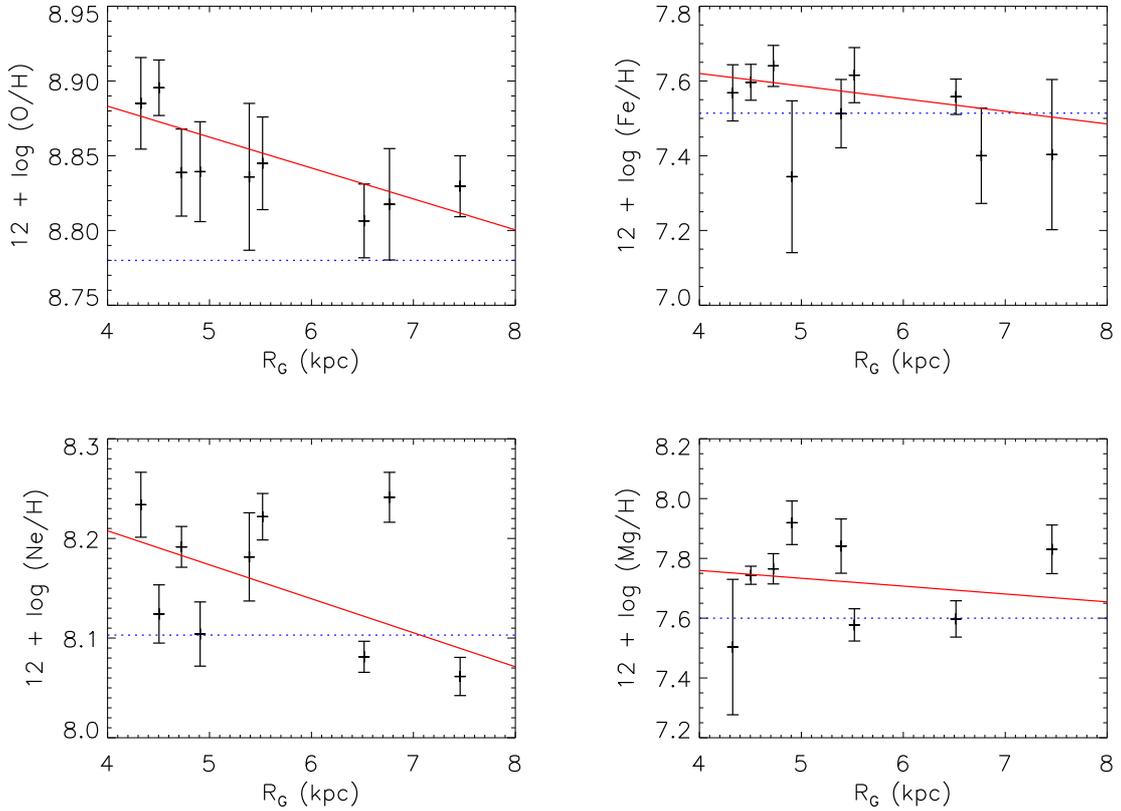}
\caption{Abundance gradients on a logarithmic scale. The blue dotted lines show the proto-Solar values.}
    \label{fig:gradients_log}
   \end{figure*}

We have also estimated the total abundances of O, Ne, Mg, and Fe for the cold phase in units of \citet{Lodders09} by summing, for each element, the contributions from neutral gas, dust, and molecules (see Table \ref{table:fit_abundances}). In most cases all the elements appear to be over-abundant with respect to the proto-Solar values. This is expected as all LOS cross inner Galactic regions where the abundances are higher. It is very difficult to estimate the location of the absorbing medium for several reasons. At first, we integrate over the LOS and thus our abundances have to be considered as large-scale properties of the ISM. Moreover, the ISM mass-density distribution within the Galaxy is not well known and the LMXB distances are affected by a large uncertainty which will dominate the error on the ISM absorber location. Therefore, we adopt half of the distance to each LMXB as a first-order, average-distance indicator. This means that the Galactocentric radius $R_{\rm G}$ of each ISM absorber will be 
defined as the distance from the Galactic Center (hereafter GC) to the mean point of the segment which connects the Sun to each target. We show the abundance trends for O, Ne, Mg, and Fe with respect to $R_{\rm G}$ (in kpc) in Fig.~\ref{fig:gradients_log}. These trends were fitted with a linear function and we obtained the following abundance gradients: 
\begin{eqnarray}{}
  12 + \log\,({\rm O/H})  &=& (8.97 \pm 0.04) - (0.021 \pm 0.008) \times R_{\rm G}\,, \nonumber \\
  12 + \log\,({\rm Fe/H}) &=& (7.76 \pm 0.14) - (0.034 \pm 0.026) \times R_{\rm G}\,, \nonumber \\
  12 + \log\,({\rm Ne/H}) &=& (8.34 \pm 0.04) - (0.034 \pm 0.007) \times R_{\rm G}\,, \nonumber \\
  12 + \log\,({\rm Mg/H}) &=& (7.87 \pm 0.12) - (0.026 \pm 0.023) \times R_{\rm G}\,. \nonumber 
\end{eqnarray}
The plots show that only the oxygen abundance is clearly anti-correlated with the distance from the GC. {Neon also shows a gradient but with a large scatter probably due to contribution intrinsic to the sources (see e.g. \citealt{Juett2003} and \citealt{Madej2010}).} However, the linear fits show negative gradients for all the abundances which is consistent with the fact that stellar evolution proceeds faster in the Galactic inner regions. Interestingly, our abundance gradients agree with those estimated with different methods and in different wavelength regimes \citep{Maciel1999, Rolleston2000, Esteban05, GradPedicelli}. On average our values are slightly smaller than those found in the literature. An explanation may be that we overestimated the distance of the absorbers by a (small) systematic factor. In fact, the LMXBs are not exactly in the Galactic Plane, but are above or below by some hundreds parsec. This means that the distance of each ISM absorber, as projected on the Galactic plane, should be 
smaller than what we have adopted. The altitudes of the LMXBs also affect the abundance estimates because of the (negative) vertical metallicity gradient \citep[see e.g.][]{GradPedicelli}. The slopes of the radial and vertical gradients are similar \citep[we have compared their effects in][]{Pinto2012}. The sources have low latitudes (see Table\ref{table:1}), on average we obtained $\bar{b}\sim6.1^{\circ}$, which implies that the radial abundance variations are about 10 times larger than the vertical ones. This explains why ignoring the vertical gradient may only marginally affect our results. {We may have underestimated the Mg and Fe abundance gradients also because in the cold phase these elements are locked up into dust. In fact, as reported in Sect.~\ref{sec:limitations}, dust could be found at distances shorter than the gas along our LOS. This means that the average distances of the Fe and Mg absorbers may be smaller and the gradients higher than those estimated.} However, once taken all this into 
account, our estimates of abundance gradients agree even better with the literature.

We can also predict the local proto-Solar interstellar abundances and check if they agree with those found in the Solar system. According to the equation above, we calculate that at the Galactocentric radius of the Sun, which is $\sim$8.0\,kpc \citep[see e.g.][]{Malkin2012}, the predicted proto-Solar abundances are
\begin{eqnarray}{}
  12 + \log\,({\rm O/H})  _{\bigodot} &=& (8.80 \pm 0.08)\,, \nonumber   \\
  12 + \log\,({\rm Fe/H}) _{\bigodot} &=& (7.48 \pm 0.25)\,, \nonumber   \\
  12 + \log\,({\rm Ne/H}) _{\bigodot} &=& (8.07 \pm 0.07)\,, \nonumber   \\
  12 + \log\,({\rm Mg/H}) _{\bigodot} &=& (7.66 \pm 0.22)\,. \nonumber   
\end{eqnarray}
These values agree very well with the abundances in the Solar System as measured by \citet{Lodders09} [O/H]$=8.76$, [Fe/H]$=7.54$, [Ne/H]$=7.95$, [Mg/H]$=7.62$, and further validate our method. {Interestingly, the interstellar oxygen abundance, as measured by previous work, is always lower than the abundance estimates for the Solar System, OB stars, \ion{H}{ii} regions, planetary nebulae, Cepheids, and Sun-like stars (see e.g. \citealt{Stasiska2012} and \citealt{Nieva2012}). Instead, our measurements and predictions agree very well with the abundances of these stellar indicators, which confirms that the local Galaxy is chemically homogeneous (see also \citealt{Stasiska2012}).}

As clearly seen in Fig.~\ref{fig:gradients_log}, our work is optimal to probe ISM abundances within 4--8\,kpc from the GC, but this range could be easily extended by using more sources. This may be an interesting proxy for a future work extended to more LOS and background source types like LMXBs, novae and AGN (especially Blazars, which do not show significant intrinsic features).

\begin{table*}
\renewcommand{\arraystretch}{1.1}
\caption{Oxygen diagnostics: percentage contributions from the different phases and total column density.}
\begin{center}
% use packages: array
 \small\addtolength{\tabcolsep}{-2pt}
\scalebox{1}{%
\begin{tabular}{llllllllll}
 \hline
Phases         & 4U1254--69 & 4U 1636--54 & 4U 1735--44 & Aql X--1 & GS 1826--24 & GX 339--4 & GX 9+9 & SAX J1808.4 & Ser X-1         \\                                    
\hline
 \ion{O}{i}\,{\footnotesize{(dust)}}\,$^{(a)}$ & 14   & 19   & 14   & 14   & 16   & 14   & 17   &  9   & 15 \\
 \ion{O}{i}\,{\footnotesize{(gas)}}\,$^{(a)}$  & 80   & 72   & 78   & 82   & 78   & 77   & 78   & 57   & 75 \\
 \ion{O}{ii-v}\,$^{(a)}$                       &  5   &  7   &  6   &  3   &  5   &  7   &  5   & 17   &  9 \\
 \ion{O}{vi-viii}\,$^{(a)}$                    &  1   &  2   &  2   &  1   &  1   &  2   & $<$1 & 17   &  1 \\
\hline
 \ion{O}{tot}\,$^{(b)}$                        & 2.10 & 2.59 & 2.79 & 3.31 & 2.93 & 4.53 & 1.66 & 1.37 & 3.72 \\
\hline
\end{tabular}}
\label{table:fit_colfrac}
\end{center}
$^{(a)}$ Column ratios are expressed in percentage. $^{(b)}$ Oxygen total column densities are in linear units of $\,(10^{22}$\,m$^{-2})$.\\
\end{table*}

% \begin{landscape} %LANDSCAPE BEGIN
\begin{table*}
\renewcommand{\arraystretch}{1.1}
\caption{Total abundances of the cold neutral phase (gas + dust).}
\begin{center}
% use packages: array
 \small\addtolength{\tabcolsep}{-2pt}
\scalebox{1}{%
\begin{tabular}{llllllllll}
 \hline
Species            & 4U1254--69        & 4U 1636--54       & 4U 1735--44        & Aql X--1               & GS 1826--24     & GX 339--4           & GX 9+9              & SAX J1808.4     & Ser X-1         \\                                    
\hline
 O/H$\,^{(a)}$  & 1.12 $\pm$ 0.05  & 1.16 $\pm$ 0.08 & 1.14 $\pm$ 0.09  & 1.06 $\pm$ 0.06  & 1.14 $\pm$ 0.08 & 1.30 $\pm$ 0.06  & 1.27 $\pm$ 0.09 & $1.09 \pm 0.09$ & 1.13 $\pm$ 0.13 \\
 Ne/H$\,^{(a)}$ & 0.91 $\pm$ 0.04 &1.315 $\pm$ 0.07 & 1.00 $\pm$ 0.07 & 0.95 $\pm$ 0.03 & 1.23 $\pm$ 0.06 & 1.05 $\pm$ 0.07 & 1.35 $\pm$ 0.10 & $1.37 \pm 0.08$ & 1.20 $\pm$ 0.12 \\
 Mg/H$\,^{(a)}$ & 1.7  $\pm$ 0.4   & 0.95 $\pm$ 0.12 & 2.10 $\pm$ 0.35   & 1.00 $\pm$ 0.14  & 1.47 $\pm$ 0.17 & 1.39 $\pm$ 0.10  & 0.8 $\pm$ 0.4  & $<$\,0.3\,$^{(b)}$ & 1.75 $\pm$ 0.37 \\
 Fe/H$\,^{(a)}$ & 0.77 $\pm$ 0.36 & 1.26 $\pm$ 0.21 & 0.7  $\pm$ 0.3      & 1.11 $\pm$ 0.12  & 1.34 $\pm$ 0.17 & 1.21 $\pm$ 0.13  & 1.13 $\pm$ 0.20 & $0.77 \pm 0.23$ & 1.00 $\pm$ 0.21 \\
\hline
\end{tabular}}
\label{table:fit_abundances}
\end{center}
$^{(a)}$ Abundances ratios are {in the linear proto-Solar abundance} units of \citet{Lodders09}.\\
$^{(b)}$ For magnesium abundance we obtain only an upper limit.\\
\end{table*}
% \end{landscape} %LANDSCAPE END

\section{Conclusion}
\label{sec:conclusion}

We have presented a detailed treatment of the interstellar medium towards the nine LMXBs: \object{GS 1826$-$238}, \object{GX 9+9}, \object{GX 339$-$4}, \object{Aql X$-$1}, \object{SAX J1808.4$-$3658}, \object{Ser X$-$1}, \object{4U 1254$-$690}, \object{4U 1636$-$536}, and \object{4U 1735$-$444}. These bright X-ray sources are spread over about 100 degrees in longitude and are located near the Galactic plane, which makes them a very good workbench for a study focused on the ISM in the Galactic disk.

We probed the ISM dust composition, total abundances, and abundance gradients. We have shown that along these LOS the ISM is composed of a complex mixture of a multi-phase gas, dust, and molecules. The analysis of all ionic species of oxygen showed that the gas is mostly neutral. The column densities of the neutral gas species are in agreement with those found in the literature. The column densities that we estimated for the mildly ionized gas are generally lower because previous work did not take into account that a significant part of the ISM transmission may be provided by dust. {A significant fraction of oxygen and iron is contained in solids, but their exact chemical composition is yet to be determined}. About 15--25\% and 65--90\% of the total amount of neutral oxygen and iron, respectively, is found in dust. This result provides a solution to the problem recently raised on the oxygen depletion and total abundance in the interstellar medium. Interestingly, the dust contribution and the compounds 
mixture seem to be consistent along all the lines-of-sight (LOS). The ratios between the different ionization states are also similar between the LOS. This suggests that on a large scale the ISM is chemically homogeneous. Finally, using our X-ray data we confirm the abundance gradients in the Galaxy and the local proto-Solar abundances as derived at longer wavelengths. {Through the comparison between our estimates of oxygen abundance and those obtained by other work on nearby young stars and other indicators such as \ion{H}{ii} regions and planetary nebulae, we also confirm that the local Galaxy is chemically homogeneous.} This work shows that X-ray spectroscopy is a very powerful method to probe the ISM.

\begin{acknowledgements}
This work is based on observations obtained with XMM-Newton, an
ESA science mission with instruments and contributions directly funded by
ESA Member States and the USA (NASA). SRON is supported financially by
NWO, the Netherlands Organization for Scientific Research.
We are grateful to Jean in't Zand and Frank Verbunt for the useful discussion 
about the physics and distances of the low-mass X-ray binaries.
{We also thank referee Tim Kallman very much for his comments which significantly improved
the clarity and quality of the paper.}
\end{acknowledgements}
 
\bibliographystyle{aa}
\bibliography{bibliografia} %----> bibliografia.bib

\end{document}